\documentclass[aip,
amsmath,amssymb, 
preprint, 
]{revtex4-2}

\usepackage[british]{babel} 
\usepackage[dvipsnames, table]{xcolor}
\definecolor{mygray}{gray}{0.9}
\usepackage{caption}
\DeclareCaptionFormat{custom}{
\textbf{\footnotesize#1#2}\footnotesize#3
}
\usepackage{lipsum}

\usepackage{graphicx} 
\usepackage{tabularx}
\usepackage{multirow}
\usepackage{colortbl}
\usepackage{array}  
\usepackage{esint}
\usepackage{amsmath}
\usepackage{amssymb}
\allowdisplaybreaks
\usepackage{mathtools} 
\usepackage{physics} 
\usepackage{BOONDOX-calo} 
\usepackage{bbm} 
\usepackage[T1]{fontenc}

\begin{document}
\title{
Polarisation conversion and optical meron topologies 
in anisotropic epsilon-near-zero metamaterials
}

\author{Vittorio~Aita}
\affiliation{Department of Physics and London Centre for Nanotechnology, King's College London, Strand, London WC2R 2LS, UK}

\author{Anastasiia~Zaleska} 
\affiliation{Department of Physics and London Centre for Nanotechnology, King's College London, Strand, London WC2R 2LS, UK}

\author{Henry~J.~Putley}
\affiliation{Department of Physics and London Centre for Nanotechnology, King's College London, Strand, London WC2R 2LS, UK}
\affiliation{School of Physics and Astronomy, University of Birmingham, Edgbaston, Birmingham B15 2TT, United Kingdom}

\author{Anatoly~V.~Zayats}
\affiliation{Department of Physics and London Centre for Nanotechnology, King's College London, Strand, London WC2R 2LS, UK}

\begin{abstract}
Plasmonic metamaterials provide a flexible platform for light manipulation and polarisation management, thanks to their engineered optical properties with exotic dispersion regimes. Here, we exploit the enhanced spin-orbit coupling induced by the strong anisotropy of plasmonic nanorod metamaterials to control the polarisation of vector vortex beams and generate complex field structures with meron topology. Modifying the degree of ellipticity of the input polarisation, we show how the observed meron topology can be additionally manipulated. Flexible control of the state of polarisation of vortex beams is important in optical manipulation, communications, metrology and quantum technologies. 

\end{abstract}

\date{\today}

\maketitle

\section{Introduction}
Polarisation-controlled light-matter interactions are important in modern technologies from optical communications and sensing to photochemical transformations and quantum optics~\cite{Zhang2018,shen2019optical,Forbes2019,YangXie2022,Bouchal2022,wang2023coloured,Forbes2024}. The possibility to engineer optical beam and material properties provides an opportunity for designing how they influence each other~\cite{krasavin-radial,yuanyang}.
Complex topological structures, including polarisation and field quasiparticles of light were demonstrated in evanescent fields as well as propagating waves, exploiting interactions between spin and orbital angular momentum of light~\cite{Shen2024}.
Uniaxial materials are important in this respect as they provide optical spin-orbit coupling which can be used for the generation of vortex beams~\cite{Ciattoni2003}. Uniaxial metamaterials, on the other hand, provide a much stronger anisotropy, leading to an enhancement of both spin-orbit coupling~\cite{aita2022enhancement} and chiral response~\cite{ginzburg-opex}.  

Plasmonic nanorod-based metamaterials are known for achieving various dispersion regimes due to their epsilon-near-zero (ENZ) properties. In the spectral range of their hyperbolic dispersion, these metamaterials exhibit strong anisotropy and respond as either a metal or a dielectric to light of different polarisations~\cite{Roth2024}. The ENZ behaviour exclusively affects fields polarised along the nanorods (parallel to the optical axis of the metamaterial), so that under plane wave illumination, oblique incidence is required to access this regime.
This condition on the electric field can be achieved at normal incidence by strongly focusing either scalar or vector beams to generate a non-negligible longitudinal field component~\cite{aita2024radazi}. The combination of structured light with engineered plasmonic metamaterials can therefore be exploited to achieve strong spin-orbit coupling and to tailor the polarisation of optical fields. 

The interaction of vector beams with uniaxial media can be used to control vector vortex beams and transform their polarisation, for example, into azimuthal or complex vortex patterns, depending on the dispersion regime of the metamaterial~\cite{aita2022enhancement}. It was shown theoretically that a non-ideal radially polarised beam, \textit{i.e.} in the case of polarisation locally elliptical, develops a vorticity whose direction is mediated by the birefringence and the sign of the linear dichroism of the metamaterial through spin-orbit coupling, as well as influenced by the longitudinal field in the ENZ regime~\cite{aita2022enhancement}.   
Here, we experimentally demonstrate polarisation control of vector vortex beams with an anisotropic plasmonic metamaterial 
in its ENZ and hyperbolic regimes (Fig.~\ref{fig:system}). In the former case, we demonstrate the azimuthalisation of the input polarisation, whereas in the latter, we reveal the emergence of vortex-like polarisation structures that possess second-order meron topology.

\section{Results and discussion}
\begin{figure}[!t]
\centering
    \includegraphics[width = 0.8\textwidth]{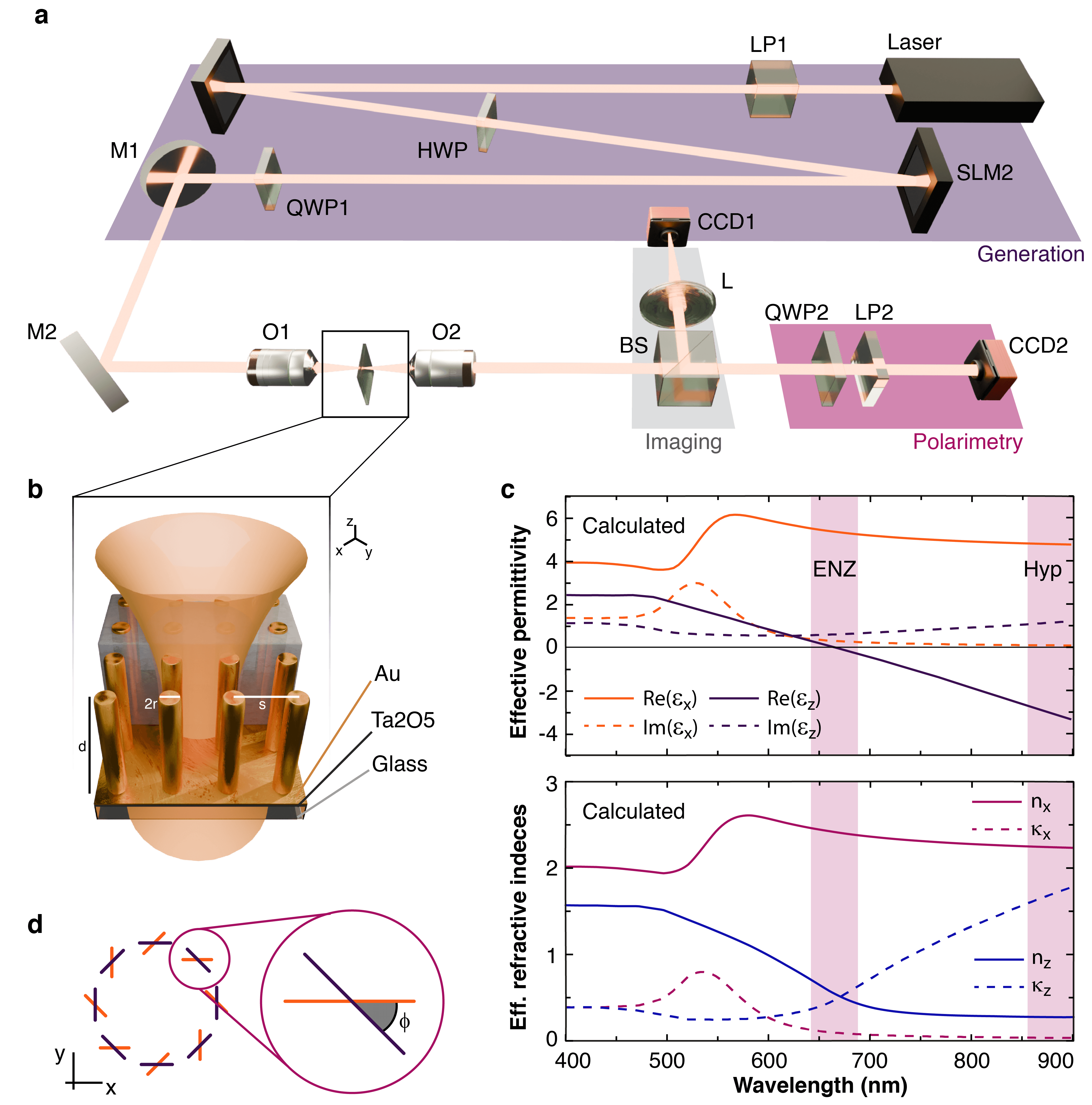}
    \caption{
    (a) Schematic of the experimental setup with (b) an illustration of the metamaterial under focused illumination. (c) Components of (top) the effective permittivity tensor and (bottom) the real (n) and imaginary ($\kappa$) parts of the ordinary (x) and extraordinary (z) refractive index of the metamaterial obtained with the Maxwell Garnett approximation (Eq.~\eqref{eq:MG}). (d) Local projection of an arbitrary state of polarisation (orange) onto the azimuthal state (purple), with the inset showing the definition of the angle $\phi$ between them.}
    \label{fig:system}
\end{figure}
Vector vortex beams are usually described as a superposition of two scalar vortices carrying a topological charge $\ell$ and orthogonal circular (left-handed (L) and right-handed (R)) polarisation states~\cite{Milione2012}, making it convenient to express them in terms of Laguerre-Gauss modes (LG$_{\ell p}$): 
\begin{equation}\label{eq:psi_general}
    \ket{\psi} = \ket{\sigma_1} \braket{\sigma_1}{\psi}\mathrm{LG}_{\ell_1,p} + \ket{\sigma_2} \braket{\sigma_2}{\psi}\mathrm{LG}_{\ell_2,p}\,,
\end{equation}
where $p$ is the radial quantum number describing the LG modes, $\ket{\sigma_i}$ represent a circular state of polarisation with spin $\sigma_i$ ($\ket{\sigma_i} = (\hat{x} -\sigma_i\hat{y})/\sqrt{2}$), and Dirac notation is used to express the projections of the state $\ket{\psi}$ onto the basis vectors $\{\ket{\mathrm{R}}, \ket{\mathrm{L}}\}$.
By fixing the spin ($\sigma_i$) and angular ($\ell_i)$ momenta of each term in Eq.~\ref{eq:psi_general}, the state $\ket{\psi}$ can be obtained as a superposition of eigenmodes of the total angular momentum $J = L + S$. The subspace of $J=0$ can be obtained for $\ell = \pm 1$ and $\sigma = \mp 1$, which span the orthogonal states describing radial ($\ket{\mathrm{Rad}}$) and azimuthal ($\ket{\mathrm{Azi}}$) polarisations:
\begin{figure}[!t]
\centering
    \includegraphics[width = 0.8\textwidth]{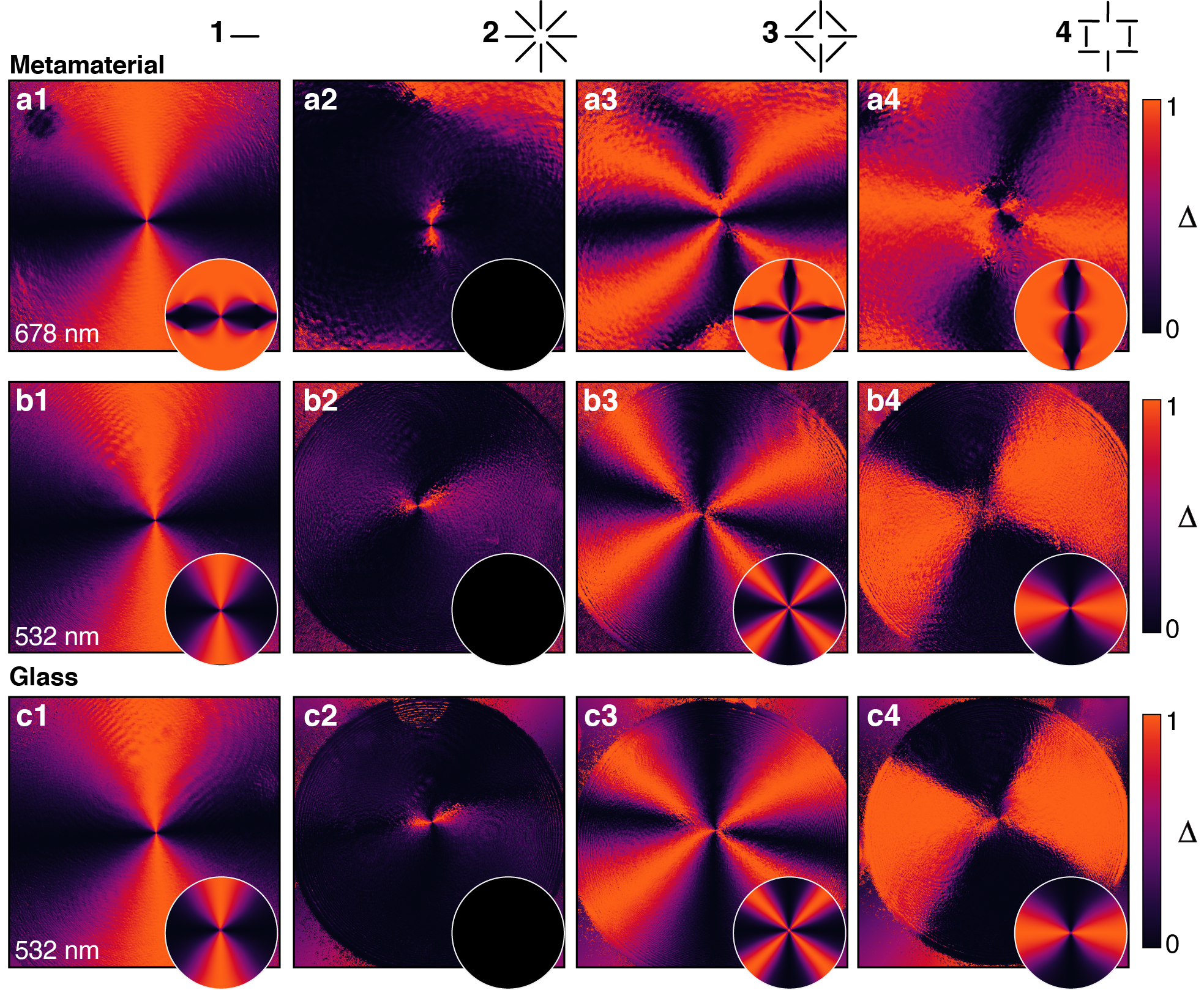}
    \caption{
    Measured projections of various states of polarisation of vector beams onto the azimuthal state: (1) linear (horizontal), (2) radial, (3) ``anti-radial'' and (4) second-order radial. The spatial maps are obtained for propagation through (a,b) the metamaterial and (c) glass, for a wavelength of (a) $\lambda_L$ = 678~nm and (b,c) $\lambda_T=532$~nm. Circular insets show simulation results for each case. The deviation from the azimuthal state is measured as $\Delta = \cos^2(\phi_\psi-\phi_\mathrm{Azi})$.}
    \label{fig:azi_proj}
\end{figure}
\begin{subequations}\label{eq:rad_azi_basis}
\begin{align}
    \ket{\mathrm{Rad}} & = \mathrm{LG}_{-10}\ket{1} + \mathrm{LG}_{10}\ket{-1}\\
    \ket{\mathrm{Azi}} & = \mathrm{LG}_{-10}\ket{1} - \mathrm{LG}_{10}\ket{-1}\,.
\end{align}
\end{subequations}
It should be noted that Eq.~\ref{eq:psi_general} does not satisfy Maxwell's equations, and the longitudinal field of an appropriate amplitude $E(z)$ should be added  
to ensure a divergence of zero ($\mathbf{\nabla}\cdot\mathbf{E}=0$)~\cite{aita2024PRB,aita2024radazi}, while the electric field of the azimuthal state is perfectly two-dimensional. 

Experimentally, we generated complex vector beams with the desired polarisation structure by employing two spatial light modulators (Fig.~\ref{fig:system}a,b, see Methods for the details). The beam is then focused at normal incidence along the optical axis of the metamaterial by an objective with numerical aperture $\mathrm{NA}=0.85$ and the transmitted light is collected with a second objective of $\mathrm{NA}=0.9$. After passing through the metamaterial, the transmitted light beam is imaged with a CCD camera and polarimetry measurements are performed (see Methods). 

The beams with linear ($\ell_{1,2} = p = 0,\, \braket{\sigma_1}{\psi} = \braket{\sigma_2}{\psi} = 1$), radial ($\ket{\mathrm{R}}$ in Eq.~\ref{eq:rad_azi_basis}a), ``anti-radial'' ($\ell_1 = \sigma_1=-\ell_2 = -\sigma_2 = 1,\,p=0,\,\braket{\sigma_1}{\psi} = \braket{\sigma_2}{\psi} = 1$) and second-order radial ($\ell_1 = -\ell_2=-2,\,\sigma_1 = -\sigma_2 = 1,\,p=0,\,\braket{\sigma_1}{\psi} = \braket{\sigma_2}{\psi} = 1$) polarisation structures were investigated (Fig.~\ref{fig:azi_proj}). In addition to pure radially polarised beams, which are characterised by a transverse spin, several vector beams were studied with a modified radial polarisation by introducing an increasing degree of ellipticity in the transverse polarisation, thus introducing also a longitudinal spin component.  

\subsection{ENZ regime: polarisation azimuthalisation.}
In the epsilon-near-zero regime of the metamaterial ($\lambda_{\rm ENZ}\approx$ 680~nm), the strong damping of the longitudinal field causes the two-dimensional polarisation to re-arrange into an azimuthal state (``azimuthalisation'') for all studied beams but the pure radial one (Fig.~\ref{fig:azi_proj}a,b). On the contrary, the polarisation is left perfectly unchanged for propagation at a wavelength far from the ENZ regime ($\lambda_{\mathrm{T}}\approx$ 530~nm) or through simple glass~(Fig.~\ref{fig:azi_proj}c). The changes occurring in the state of polarisation are calculated from the experimental and simulated polarisation distributions as a deviation of the polarisation angle of the beam ($\phi_\psi$) from that of an azimuthal beam ($\phi_\mathrm{Azi}$) as $\Delta = \cos^2\phi = \cos^2(\phi_\psi-\phi_\mathrm{Azi})$ (Fig.~\ref{fig:system}d). Experimental observations and theoretical predictions are in a good agreement, with almost complete azimuthalisation of the input polarisation. The observed differences can be ascribed to imperfections in the experimentally generated polarisation states which, although globally reproducing the symmetry of the desired vector beams, suffer from a remaining non-zero local ellipticity. The main consequence of this non-zero ellipticity is a reduction of the longitudinal field strength which causes a drastically lower coupling to the ENZ response of the metamaterial~\cite{aita2022enhancement}, diminishing the development of azimuthalisaiton in the experiment. This can also be understood from the point of view of reduction of the transverse spin and dominating longitudinal spin. This process is restricted in pure radially polarised beams due to the requirements of zero-divergence electric field~\cite{aita2024radazi}. For comparison, propagation through glass does not result in the azimuthalisation, as expected, since glass does not influence the balance between transverse and longitudinal field components.  
\begin{figure}[!t]
\centering
    \includegraphics[width =\textwidth]{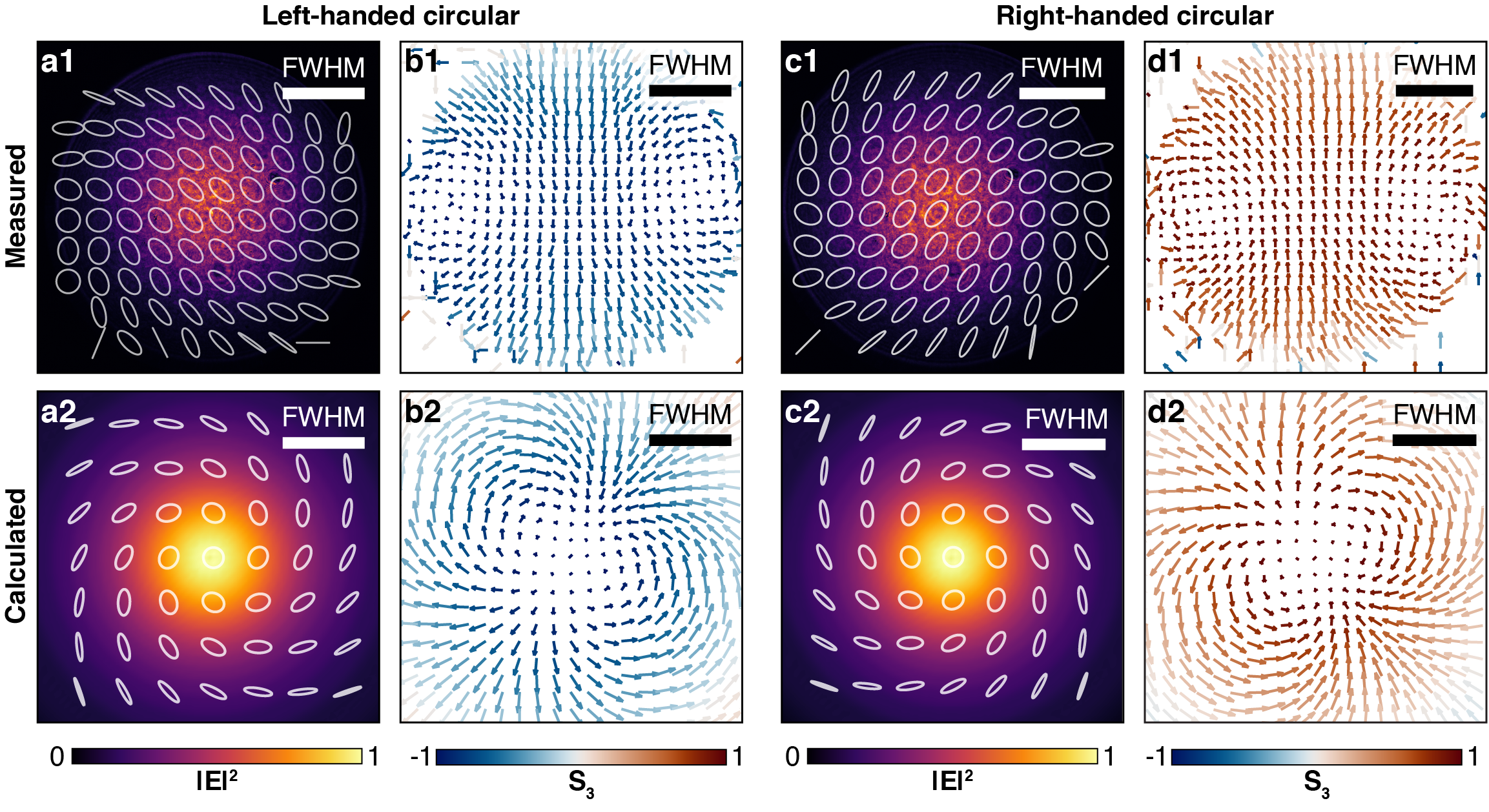}
    \caption{\textbf{Second-order meron topology.} (1) Experimental and (2) theoretical results for the (left) left- and (right) right-handed circularly polarised Gaussian beam, tightly focused ($\mathrm{NA} = 0.85$) through the metamaterial: (a,c) state of polarisation recovered from the Stokes parameters (see Methods) overlapped with the intensity profile of the beam, (b,d) representation of the vector field $\vb{\Sigma}$ in the $x-y$ plane. The colour of the arrows depicts the $z$-component of the field.}
    \label{fig:off_res}
\end{figure}
\newline\\
\subsection{Hyperbolic regime: generation of second-order meron topologies}
In the hyperbolic dispersion regime, the plasmonic nanorod metamaterial offers strong anisotropy, observed at wavelengths longer than $\lambda_{\mathrm{ENZ}}$ (Fig.~\ref{fig:system}c). Both strong birefringence and strong dichroism are present in this spectral range. The spin-orbit coupling enabled by elliptical polarisation---and enhanced by the tight focusing---can be exploited 
in order to realise vortex polarisation structures after the transmission through the metamaterials~\cite{Ciattoni2003,aita2022enhancement}. Circularly polarised beams propagating through an array of nanorods away from the ENZ regime experience a strong modification of their polarisation state and vorticity (longitudinal spin and orbital angular momentum.). The output polarisation shows a non-uniform spatial distribution (Fig.~\ref{fig:off_res}a,c) with the ellipticity changing with the distance from the beam centre, and the orientation of the local polarisation creating a vortex structure whose global orientation depends on the sign of the helicity of the initial state ($\sigma = \pm1$). 

The origin of this state of polarisation is found in the interplay between the anisotropy offered by the metamaterial and the tight focusing of the incoming circular beam. Propagating along the optical axis of a uniaxial material, a circularly polarised beam with a circularly symmetric intensity profile generates an optical vortex of order two with the conversion efficiency increasing with the focusing (Supplementary Fig.~\ref{fig:si_circ_vortex_comps}). This is realised in the circular component of the transmitted light with opposite spin to the input~\cite{Ciattoni2003}, so that the angular momentum is conserved (\textit{e.g.}, an input of $\sigma = 1,\ell=0$  produces the vortex with $(\sigma=-1,\ell=2)$ in the output). The superposition of these two components carrying different orbital angular momenta (2 and 0) and having orthogonal circular polarisations creates the state of polarisation observed here. Remarkably, although the metamaterial used is considerably thin---the rod height is approximately 250~nm ($<<\lambda$)---the vortex component generated is strong enough to modify the input circular polarisation. This is achieved thanks to the strong anisotropy offered by the metamaterial ($\Delta n = |n_x - n_z| \approx 1.8$, see Fig.~\ref{fig:system}c) and to the tight focusing (see Supplementary Fig.~\ref{fig:si_circ_vortex_comps}), which increases the conversion rate of the input circularly polarised beam to the vortex beam of the orthogonal polarization, making it comparable to the stronger circular one, resulting in the observed polarisation distribution. 

The obtained structure can be described by the spatial distribution of a Cartesian vector ($\vb{\Sigma}$) of components given by the Stokes parameters $(S_1,S_2,S_3)$~\cite{Gao2020}, normalised to obtain a unit vector at every point $(x,y)$. This reveals the emergence of a synthetic topological structure in the polarisation of the beam (Fig.~\ref{fig:off_res}b,d). The skyrmion number $\mathrm{SN}$ of this structure can be computed as~\cite{Shen2024}
\begin{equation}
    \mathrm{SN} = \frac{1}{4\pi}\iint_\Omega\vb{\Sigma}\cdot\qty(\pderivative{x}\vb{\Sigma}\cross\pderivative{y}\vb{\Sigma})\,\dd x\dd y\,,
\end{equation}
integrated over the $x-y$ plane perpendicular to the propagation direction ($\hat{z}$). From the numerically simulated polarisation patterns, using as the integration domain $\Omega$ a circle with diameter equal to the beam full-width half-maximum (FWHM), we can obtain a skyrmion number $\mathrm{SN}\approx \pm1.05$ ~(Supplementary Figs.~\ref{fig:si_LHCpoin} and \ref{fig:si_RHCpoin}), with sign dependent on the choice of initial helicity. 
Although a unitary SN would suggest the generation of a skyrmion of order one, the obtained topology corresponds to only partial coverage of the Poincaré sphere~(see SI Figs.~\ref{fig:si_LHCpoin},\ref{fig:si_RHCpoin}). The vector field $\vb{\Sigma}$ only covers one of the hemispheres, depending on the sign of the initial helicity. This observation together with the spatial distribution of $\vb{\Sigma}$~(Fig.~\ref{fig:off_res}b,d) rather suggests that a second-order Stokes meron is observed~\cite{Krol2021}. Conversely, there would be full coverage of the Poincaré sphere if it were a bi-meron topology, which consists of two merons of opposite signs~\cite{Shen2021}. The half-coverage shown by our results could alternatively be achieved by a meron pair, although in this case the vortex points of the two merons should be distinguishable~\cite{Dreher2024}. The topological texture in the normalised Stokes vector $\vb{\Sigma}$ can be visualised as two joined merons with the same vorticity ($\pm1/2$) such that the unitary skyrmion number is explained as the sum of two half-integers with the same sign. 

Polarimetry measurements performed on tightly focused ($\mathrm{NA} =$0.85) circularly polarised Gaussian beams transmitted through the nanorod metamaterial ($\lambda \approx$ 800~nm) reproduce a state of polarisation with a similar structure to what is predicted by simulations (Fig.~\ref{fig:off_res}). The experimental reconstruction of the vector field $\vb{\Sigma}$ for right- (left)-handed input also shows an always positive (negative) $\Sigma_z$, with a negligible presence of data points in the southern (northern) hemisphere of the Poincaré sphere~(see Supplementary Figs.~\ref{fig:si_LHCpoin} and\ref{fig:si_RHCpoin}). This results in a topological structure that does not quite reproduce the second-order meron predicted from calculations, but rather two merons of the same vorticity that are not yet joined~\cite{Krol2021}. It should be noted that small disorder either in the metamaterial structure or the incident beam may result in the splitting of vortices leading to this observation~\cite{afanasiev-APN}. This translates to a non-unitary skyrmion number that also fluctuates considerably with the spatial limits chosen for the integration domain $\Omega$. The realisation of a second-order meron topology is enabled by the spin-orbit coupling, enhanced by the strong anisotropy of the metamaterial. The efficiency of this process depends on two factors: the degree of anisotropy of the metamaterial, and the spin angular momentum density of the initial state of polarisation.

Calculations and measurements performed at wavelengths lower than $\lambda_{\mathrm{ENZ}}$ show that the anisotropy provided by the nanorod metamaterial in this range is considerably weaker than what is obtained with hyperbolic dispersion. The theoretical results show an almost negligible reduction of the local ellipticity (Supplementary Fig.~\ref{fig:si_blue_shift}), as well as the impossibility of re-creating the second-order meron. The Stokes parameters obtained in this case offer considerably limited coverage of the northern hemisphere of the Poincaré sphere, which is translated into a vector field $\vb{\Sigma}$ that--although seemingly reproducing a double-meron symmetry--does not cover the full range of values needed for its $z$-component, resulting in a skyrmion number of $\mathrm{SN} = 0.17$. Accordingly, the topology of interest is also lost experimentally when moving to the elliptic dispersion regime.

\begin{figure}[!t]
\centering
    \includegraphics[width = \textwidth]{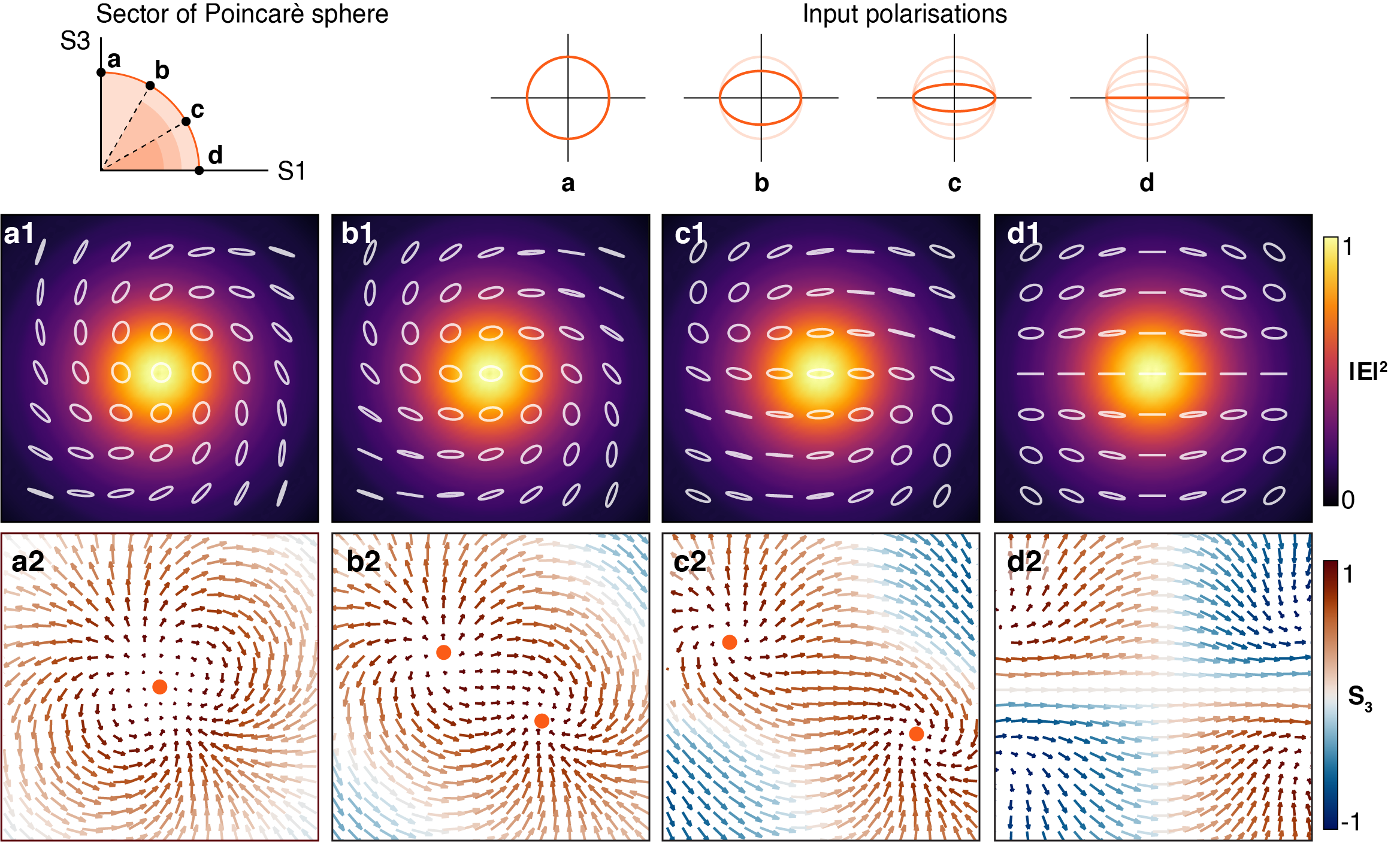}
    \caption{
    The influence of the degree of ellipticity in the initial polarisation on the second-order meron topology in the output field. (a--d) The polarisation changes from (a) circular to (d) linear as represented on the Poincaré sphere on the top-left insert: (1) Intensity profile and state of polarisation obtained after tight focusing through the metamaterial, (2) corresponding distribution of the vector field $\vb{\Sigma}$. The second-order meron in (a2) is shown to break up into single merons that drift apart as the ellipticity approaches zero.}
    \label{fig:imperfections}
\end{figure}

Lastly, a gradual reduction of the ellipticity of the input beam from fully circular ($\sigma_i = \pm 1$) polarisation is shown to make the second-order meron split into individual merons with the same vorticity, which drift apart with the deacrease of the ellipticity and eventually vanish for perfectly linear ($\sigma_i =0$) polarisation (Fig.~\ref{fig:imperfections} and Supplementary Fig.~S5). Experimentally, the beam possesses polarisation imperfections primarily affecting the local ellipticity, which can explain the complexity of achieving a fully formed second-order meron, even when the metamaterial anisotropy is strong enough to allow for its formation.

\section{Conclusions}
We have studied the interaction of vector beams carrying longitudinal field with a strongly anisotropic metamaterial and the related spin-orbit coupling effects.  
Depending on the dispersion regime of the interaction, the metamaterial has been shown to (i) modify the beam polarisation into an azimuthal state (ENZ regime) or (ii) generate a vortex-like structure in polarisation with second-order meron topology, which strongly depends on the metamaterial anisotropy and the spin composition of the incident beam. While the strong anisotropy offered by the hyperbolic dispersion regime leads to the realisation of a second-order meron topology, the weaker anisotropy characteristic of the elliptic dispersion produces a topologically trivial structure.
Experimental results consistently reproduce the theoretical predictions taking into account imperfections in the local beam ellipticity that has been theoretically proven to drastically decrease the strength of the longitudinal field generated upon focusing. 

Previously observed quasiparticles of light in the Stokes field have been generated from superpositions of optical vortices collectively possessing a non-zero total angular momentum~\cite{Gao2020}. By tightly focusing a laser beam in a strongly anisotropic hyperbolic metamaterial, we observe the generation of Stokes merons from a simple circularly polarised beam modified by the uniaxial material. This underlines the potential of strongly anisotropic plasmonic metamaterials as a platform for beam and polarisation shaping, as well as to control the topology of optical fields.

\section{Methods}
\subsection{Metamaterial fabrication and characterisation}
The metamaterial was fabricated by electrochemical approach as described in Ref.~\cite{zaleska2024copper}. The targeted parameters of the nanostructure were the radius of the individual rods $r=16\pm1.8$~nm, the spacing (centre-to-centre) in between adjacent rods $s=60.7\pm4.8$~nm and the overall thickness of the sample $d$ in the range of 200--250 nm. 

\subsection{Semi-analytical modelling}
To model the propagation of focused laser beams through the anisotropic metamaterial, we have used a previously developed extension of the Richards-Wolf theory for anisotropic media~\cite{aita2024PRB}. The first and last layers are considered to be free space ($\varepsilon = \mu = 1$) and glass ($\varepsilon = 2.25$, $\mu = 1$), respectively (Fig.~\ref{fig:system}b). The middle layer represents the metamaterial as a bulk uniaxial crystal with the optical properties described by an effective medium theory, which models the gold nanorods as inclusions in a host alumina medium~\cite{Roth2024}. Combining tabulated data for both materials~\cite{JohnsonChristy1972} and taking into account the corrections for quality of the electrochemical gold\cite{jingyi-apl}, the non-zero components of the effective permittivity tensor are obtained as 
\begin{subequations}\label{eq:MG}
    \begin{gather}
        \varepsilon_x = \varepsilon_{\mathrm{Al_2O_3}} \frac{(1-f)\varepsilon_{\mathrm{Al_2O_3}}+(1+f)\varepsilon_\mathrm{Au}}{(1-f)\varepsilon_\mathrm{Au} + (1+f)\varepsilon_{\mathrm{Al_2O_3}}}\, ,\\
    \varepsilon_z = (1-f)\varepsilon_{\mathrm{Al_2O_3}} + f\varepsilon_\mathrm{Au}\, ,\\
   \text{where}\quad\varepsilon_\mathrm{Au} = \varepsilon_b+\frac{i\omega_P\tau\qty(\mathrm{R}_b-\mathrm{R})}{\omega\qty(\omega\tau+i)\qty(\omega\tau\mathrm{R}+i\mathrm{R}_b)}.
    \end{gather}
\end{subequations}
Here $f$ represents the filling fraction of the gold inclusions in the alumina matrix, the subscript $b$ refers to quantities characterising bulk gold, $\rm{R}$ is the electron mean free path for gold, $\omega_P$ its plasma frequency and $\tau$ the average electron collision time.

\subsection{Vectorial vortex beam generation}
The setup for beam shaping (Fig.~\ref{fig:system}a) is based on two reflective spatial light modulators (SLM) (HOLOEYE PLUTO-02 with a NIRO-023 head). Once the wavelength is selected from the supercontinuum source (Fianium Supercontinuum Femtopower1060 SC450-2) with a combination of filters, the beam is expanded with a pair of short converging lenses placed in a $4-f$ configuration ($f_1$ = 35~mm, $f_2$ = 70~mm). This allows for the possibility of spatially filtering the beam before modulation. The reference polarisation ($\ket{\mathrm{H}}$) is fixed by $\mathrm{LP}_1$ (a Glenn-Taylor prism) so that it is aligned with the horizontal axis of the SLM to maximise the modulation efficiency. The SLMs apply phase masks that apply different topological charges ($\ell_1,\,\ell_2$) while polarisation is rotated by an HWP to a diagonal state in between the two modulators. Using a QWP the co-propagating vortices are made circularly polarised, so that their superposition returns the desired vectorial state (Eq.~\ref{eq:rad_azi_basis}). By choosing the orientation of the HWP and the QWP and the values of $\ell_1,|ell_2$, the output polarisation state can be tuned, allowing for the generation of any scalar or vector beam.

\subsection{Polarimetry measurements}
A full characterisation of the polarisation state of the transmitted light was performed by adding a linear polariser and a quarter waveplate at the end of the transmission path (Fig.~\ref{fig:system}a, pink area). The unknown polarisation state is projected onto horizontal, vertical, diagonal, antidiagonal, right- and left-handed circular sates, so as to retrieve the Stokes parameters:
\begin{subequations}\label{eq:stokes_intensity}
    \begin{align}
        \mathcal{s}_0 &= I_H + I_V                  \\
        S_1 &= \Bigl(\mathrm{I_H} - \mathrm{I_V}\Bigr)/\mathcal{s}_0  \\
        S_2 &= \Bigl(\mathrm{I_D} - \mathrm{I_A}\Bigr)/\mathcal{s}_0  \\
        S_3 &= \Bigl(\mathrm{I_L} - \mathrm{I_R}\Bigr)/\mathcal{s}_0\,. 
        \end{align}
\end{subequations}
The above quantities are obtained as functions of the coordinates in the transverse plane, so that the geometrical parameters of the polarisation ellipse can be calculated for each pixel of the image, and the local polarisation fully characterised.

\section*{Acknowledgments.} This work was supported in part by the the UK EPSRC project EP/Y015673/1 and the ERC iCOMM Project (No. 789340). All the data supporting findings of this work are presented in the Results section and are available from the corresponding author upon reasonable request.

\bibliography{References.bib}

\begin{thebibliography}{25}%
\makeatletter
\providecommand \@ifxundefined [1]{%
 \@ifx{#1\undefined}
}%
\providecommand \@ifnum [1]{%
 \ifnum #1\expandafter \@firstoftwo
 \else \expandafter \@secondoftwo
 \fi
}%
\providecommand \@ifx [1]{%
 \ifx #1\expandafter \@firstoftwo
 \else \expandafter \@secondoftwo
 \fi
}%
\providecommand \natexlab [1]{#1}%
\providecommand \enquote  [1]{``#1''}%
\providecommand \bibnamefont  [1]{#1}%
\providecommand \bibfnamefont [1]{#1}%
\providecommand \citenamefont [1]{#1}%
\providecommand \href@noop [0]{\@secondoftwo}%
\providecommand \href [0]{\begingroup \@sanitize@url \@href}%
\providecommand \@href[1]{\@@startlink{#1}\@@href}%
\providecommand \@@href[1]{\endgroup#1\@@endlink}%
\providecommand \@sanitize@url [0]{\catcode `\\12\catcode `\$12\catcode `\&12\catcode `\#12\catcode `\^12\catcode `\_12\catcode `\%12\relax}%
\providecommand \@@startlink[1]{}%
\providecommand \@@endlink[0]{}%
\providecommand \url  [0]{\begingroup\@sanitize@url \@url }%
\providecommand \@url [1]{\endgroup\@href {#1}{\urlprefix }}%
\providecommand \urlprefix  [0]{URL }%
\providecommand \Eprint [0]{\href }%
\providecommand \doibase [0]{https://doi.org/}%
\providecommand \selectlanguage [0]{\@gobble}%
\providecommand \bibinfo  [0]{\@secondoftwo}%
\providecommand \bibfield  [0]{\@secondoftwo}%
\providecommand \translation [1]{[#1]}%
\providecommand \BibitemOpen [0]{}%
\providecommand \bibitemStop [0]{}%
\providecommand \bibitemNoStop [0]{.\EOS\space}%
\providecommand \EOS [0]{\spacefactor3000\relax}%
\providecommand \BibitemShut  [1]{\csname bibitem#1\endcsname}%
\let\auto@bib@innerbib\@empty
\bibitem [{\citenamefont {Zhang}\ \emph {et~al.}(2018)\citenamefont {Zhang}, \citenamefont {Liu}, \citenamefont {Gao},\ and\ \citenamefont {Yang}}]{Zhang2018}%
  \BibitemOpen
  \bibfield  {author} {\bibinfo {author} {\bibfnamefont {Y.}~\bibnamefont {Zhang}}, \bibinfo {author} {\bibfnamefont {W.}~\bibnamefont {Liu}}, \bibinfo {author} {\bibfnamefont {J.}~\bibnamefont {Gao}},\ and\ \bibinfo {author} {\bibfnamefont {X.}~\bibnamefont {Yang}},\ }\bibfield  {title} {\enquote {\bibinfo {title} {Generating focused 3d perfect vortex beams by plasmonic metasurfaces},}\ }\href@noop {} {\bibfield  {journal} {\bibinfo  {journal} {Advanced Optical Materials}\ }\textbf {\bibinfo {volume} {6}},\ \bibinfo {pages} {1701228} (\bibinfo {year} {2018})}\BibitemShut {NoStop}%
\bibitem [{\citenamefont {Shen}\ \emph {et~al.}(2019)\citenamefont {Shen}, \citenamefont {Wang}, \citenamefont {Xie}, \citenamefont {Min}, \citenamefont {Fu}, \citenamefont {Liu}, \citenamefont {Gong},\ and\ \citenamefont {Yuan}}]{shen2019optical}%
  \BibitemOpen
  \bibfield  {author} {\bibinfo {author} {\bibfnamefont {Y.}~\bibnamefont {Shen}}, \bibinfo {author} {\bibfnamefont {X.}~\bibnamefont {Wang}}, \bibinfo {author} {\bibfnamefont {Z.}~\bibnamefont {Xie}}, \bibinfo {author} {\bibfnamefont {C.}~\bibnamefont {Min}}, \bibinfo {author} {\bibfnamefont {X.}~\bibnamefont {Fu}}, \bibinfo {author} {\bibfnamefont {Q.}~\bibnamefont {Liu}}, \bibinfo {author} {\bibfnamefont {M.}~\bibnamefont {Gong}},\ and\ \bibinfo {author} {\bibfnamefont {X.}~\bibnamefont {Yuan}},\ }\bibfield  {title} {\enquote {\bibinfo {title} {Optical vortices 30 years on: Oam manipulation from topological charge to multiple singularities},}\ }\href@noop {} {\bibfield  {journal} {\bibinfo  {journal} {Light: Science \& Applications}\ }\textbf {\bibinfo {volume} {8}},\ \bibinfo {pages} {90} (\bibinfo {year} {2019})}\BibitemShut {NoStop}%
\bibitem [{\citenamefont {Forbes}\ and\ \citenamefont {Andrews}(2019)}]{Forbes2019}%
  \BibitemOpen
  \bibfield  {author} {\bibinfo {author} {\bibfnamefont {K.~A.}\ \bibnamefont {Forbes}}\ and\ \bibinfo {author} {\bibfnamefont {D.~L.}\ \bibnamefont {Andrews}},\ }\bibfield  {title} {\enquote {\bibinfo {title} {Enhanced optical activity using the orbital angular momentum of structured light},}\ }\href {https://doi.org/10.1103/PhysRevResearch.1.033080} {\bibfield  {journal} {\bibinfo  {journal} {Phys. Rev. Res.}\ }\textbf {\bibinfo {volume} {1}},\ \bibinfo {pages} {033080} (\bibinfo {year} {2019})}\BibitemShut {NoStop}%
\bibitem [{\citenamefont {Yang}\ \emph {et~al.}(2022)\citenamefont {Yang}, \citenamefont {Xie}, \citenamefont {Zhang}, \citenamefont {Ouyang}, \citenamefont {Xu}, \citenamefont {Cao}, \citenamefont {Wang}, \citenamefont {Zhu},\ and\ \citenamefont {Li}}]{YangXie2022}%
  \BibitemOpen
  \bibfield  {author} {\bibinfo {author} {\bibfnamefont {Q.}~\bibnamefont {Yang}}, \bibinfo {author} {\bibfnamefont {Z.}~\bibnamefont {Xie}}, \bibinfo {author} {\bibfnamefont {M.}~\bibnamefont {Zhang}}, \bibinfo {author} {\bibfnamefont {X.}~\bibnamefont {Ouyang}}, \bibinfo {author} {\bibfnamefont {Y.}~\bibnamefont {Xu}}, \bibinfo {author} {\bibfnamefont {Y.}~\bibnamefont {Cao}}, \bibinfo {author} {\bibfnamefont {S.}~\bibnamefont {Wang}}, \bibinfo {author} {\bibfnamefont {L.}~\bibnamefont {Zhu}},\ and\ \bibinfo {author} {\bibfnamefont {X.}~\bibnamefont {Li}},\ }\bibfield  {title} {\enquote {\bibinfo {title} {Ultra-secure optical encryption based on tightly focused perfect optical vortex beams},}\ }\href {https://doi.org/doi:10.1515/nanoph-2021-0786} {\bibfield  {journal} {\bibinfo  {journal} {Nanophotonics}\ }\textbf {\bibinfo {volume} {11}},\ \bibinfo {pages} {1063--1070} (\bibinfo {year} {2022})}\BibitemShut {NoStop}%
\bibitem [{\citenamefont {Bouchal}\ and\ \citenamefont {Bouchal}(2022)}]{Bouchal2022}%
  \BibitemOpen
  \bibfield  {author} {\bibinfo {author} {\bibfnamefont {P.}~\bibnamefont {Bouchal}}\ and\ \bibinfo {author} {\bibfnamefont {Z.}~\bibnamefont {Bouchal}},\ }\bibfield  {title} {\enquote {\bibinfo {title} {Twisted rainbow light and nature-inspired generation of vector vortex beams},}\ }\href@noop {} {\bibfield  {journal} {\bibinfo  {journal} {Laser \& Photonics Reviews}\ }\textbf {\bibinfo {volume} {16}},\ \bibinfo {pages} {2200080} (\bibinfo {year} {2022})}\BibitemShut {NoStop}%
\bibitem [{\citenamefont {Wang}\ \emph {et~al.}(2023)\citenamefont {Wang}, \citenamefont {Wang}, \citenamefont {Ruan}, \citenamefont {Chan}, \citenamefont {Zhang}, \citenamefont {Liu}, \citenamefont {Rezaei}, \citenamefont {Trisno}, \citenamefont {Qiu}, \citenamefont {Gu} \emph {et~al.}}]{wang2023coloured}%
  \BibitemOpen
  \bibfield  {author} {\bibinfo {author} {\bibfnamefont {H.}~\bibnamefont {Wang}}, \bibinfo {author} {\bibfnamefont {H.}~\bibnamefont {Wang}}, \bibinfo {author} {\bibfnamefont {Q.}~\bibnamefont {Ruan}}, \bibinfo {author} {\bibfnamefont {J.~Y.~E.}\ \bibnamefont {Chan}}, \bibinfo {author} {\bibfnamefont {W.}~\bibnamefont {Zhang}}, \bibinfo {author} {\bibfnamefont {H.}~\bibnamefont {Liu}}, \bibinfo {author} {\bibfnamefont {S.~D.}\ \bibnamefont {Rezaei}}, \bibinfo {author} {\bibfnamefont {J.}~\bibnamefont {Trisno}}, \bibinfo {author} {\bibfnamefont {C.-W.}\ \bibnamefont {Qiu}}, \bibinfo {author} {\bibfnamefont {M.}~\bibnamefont {Gu}}, \emph {et~al.},\ }\bibfield  {title} {\enquote {\bibinfo {title} {Coloured vortex beams with incoherent white light illumination},}\ }\href@noop {} {\bibfield  {journal} {\bibinfo  {journal} {Nature Nanotechnology}\ }\textbf {\bibinfo {volume} {18}},\ \bibinfo {pages} {264--272} (\bibinfo {year} {2023})}\BibitemShut {NoStop}%
\bibitem [{\citenamefont {Forbes}\ and\ \citenamefont {Green}(2024)}]{Forbes2024}%
  \BibitemOpen
  \bibfield  {author} {\bibinfo {author} {\bibfnamefont {K.~A.}\ \bibnamefont {Forbes}}\ and\ \bibinfo {author} {\bibfnamefont {D.}~\bibnamefont {Green}},\ }\bibfield  {title} {\enquote {\bibinfo {title} {Topological-charge-dependent dichroism and birefringence of optical vortices},}\ }\href@noop {} {\bibfield  {journal} {\bibinfo  {journal} {Laser \& Photonics Reviews}\ }\textbf {\bibinfo {volume} {18}},\ \bibinfo {pages} {2400109} (\bibinfo {year} {2024})}\BibitemShut {NoStop}%
\bibitem [{\citenamefont {Krasavin}\ \emph {et~al.}(2018)\citenamefont {Krasavin}, \citenamefont {Segovia}, \citenamefont {Dubrovka}, \citenamefont {Olivier}, \citenamefont {Wurtz}, \citenamefont {Ginzburg},\ and\ \citenamefont {Zayats}}]{krasavin-radial}%
  \BibitemOpen
  \bibfield  {author} {\bibinfo {author} {\bibfnamefont {A.}~\bibnamefont {Krasavin}}, \bibinfo {author} {\bibfnamefont {P.}~\bibnamefont {Segovia}}, \bibinfo {author} {\bibfnamefont {R.}~\bibnamefont {Dubrovka}}, \bibinfo {author} {\bibfnamefont {N.}~\bibnamefont {Olivier}}, \bibinfo {author} {\bibfnamefont {G.}~\bibnamefont {Wurtz}}, \bibinfo {author} {\bibfnamefont {P.}~\bibnamefont {Ginzburg}},\ and\ \bibinfo {author} {\bibfnamefont {A.}~\bibnamefont {Zayats}},\ }\bibfield  {title} {\enquote {\bibinfo {title} {Generalization of the optical theorem: experimental proof for radially polarized beams},}\ }\href@noop {} {\bibfield  {journal} {\bibinfo  {journal} {Light: Science \& Applications}\ }\textbf {\bibinfo {volume} {7}},\ \bibinfo {pages} {36} (\bibinfo {year} {2018})}\BibitemShut {NoStop}%
\bibitem [{\citenamefont {Xie}\ \emph {et~al.}(2025)\citenamefont {Xie}, \citenamefont {A.V}, \citenamefont {DJ},\ and\ \citenamefont {Zayats}}]{yuanyang}%
  \BibitemOpen
  \bibfield  {author} {\bibinfo {author} {\bibfnamefont {Y.}~\bibnamefont {Xie}}, \bibinfo {author} {\bibfnamefont {K.}~\bibnamefont {A.V}}, \bibinfo {author} {\bibfnamefont {R.}~\bibnamefont {DJ}},\ and\ \bibinfo {author} {\bibfnamefont {A.}~\bibnamefont {Zayats}},\ }\bibfield  {title} {\enquote {\bibinfo {title} {Unidirectional chiral scattering from single enantiomeric plasmonic nanoparticles},}\ }\href@noop {} {\bibfield  {journal} {\bibinfo  {journal} {Nature Communications}\ }\textbf {\bibinfo {volume} {16}},\ \bibinfo {pages} {NNNNN} (\bibinfo {year} {2025})}\BibitemShut {NoStop}%
\bibitem [{\citenamefont {Shen}\ \emph {et~al.}(2024)\citenamefont {Shen}, \citenamefont {Zhang}, \citenamefont {Shi}, \citenamefont {Du}, \citenamefont {Yuan},\ and\ \citenamefont {Zayats}}]{Shen2024}%
  \BibitemOpen
  \bibfield  {author} {\bibinfo {author} {\bibfnamefont {Y.}~\bibnamefont {Shen}}, \bibinfo {author} {\bibfnamefont {Q.}~\bibnamefont {Zhang}}, \bibinfo {author} {\bibfnamefont {P.}~\bibnamefont {Shi}}, \bibinfo {author} {\bibfnamefont {L.}~\bibnamefont {Du}}, \bibinfo {author} {\bibfnamefont {X.}~\bibnamefont {Yuan}},\ and\ \bibinfo {author} {\bibfnamefont {A.~V.}\ \bibnamefont {Zayats}},\ }\bibfield  {title} {\enquote {\bibinfo {title} {Optical skyrmions and other topological quasiparticles of light},}\ }\href {https://doi.org/10.1038/s41566-023-01325-7} {\bibfield  {journal} {\bibinfo  {journal} {Nature Photonics}\ }\textbf {\bibinfo {volume} {18}},\ \bibinfo {pages} {15--25} (\bibinfo {year} {2024})}\BibitemShut {NoStop}%
\bibitem [{\citenamefont {Ciattoni}, \citenamefont {Cincotti},\ and\ \citenamefont {Palma}(2003)}]{Ciattoni2003}%
  \BibitemOpen
  \bibfield  {author} {\bibinfo {author} {\bibfnamefont {A.}~\bibnamefont {Ciattoni}}, \bibinfo {author} {\bibfnamefont {G.}~\bibnamefont {Cincotti}},\ and\ \bibinfo {author} {\bibfnamefont {C.}~\bibnamefont {Palma}},\ }\bibfield  {title} {\enquote {\bibinfo {title} {Circularly polarized beams and vortex generation in uniaxial media},}\ }\href {https://doi.org/10.1364/JOSAA.20.000163} {\bibfield  {journal} {\bibinfo  {journal} {J. Opt. Soc. Am. A}\ }\textbf {\bibinfo {volume} {20}},\ \bibinfo {pages} {163--171} (\bibinfo {year} {2003})}\BibitemShut {NoStop}%
\bibitem [{\citenamefont {Aita}\ and\ \citenamefont {Zayats}(2022)}]{aita2022enhancement}%
  \BibitemOpen
  \bibfield  {author} {\bibinfo {author} {\bibfnamefont {V.}~\bibnamefont {Aita}}\ and\ \bibinfo {author} {\bibfnamefont {A.~V.}\ \bibnamefont {Zayats}},\ }\bibfield  {title} {\enquote {\bibinfo {title} {Enhancement of optical spin-orbit coupling in anisotropic enz metamaterials},}\ }\href@noop {} {\bibfield  {journal} {\bibinfo  {journal} {IEEE Photonics Journal}\ }\textbf {\bibinfo {volume} {15}},\ \bibinfo {pages} {1--8} (\bibinfo {year} {2022})}\BibitemShut {NoStop}%
\bibitem [{\citenamefont {Vestler}\ \emph {et~al.}(2018)\citenamefont {Vestler}, \citenamefont {Shishkin}, \citenamefont {Gurvitz}, \citenamefont {Nasir}, \citenamefont {Ben-Moshe}, \citenamefont {Slobozhanyuk}, \citenamefont {Krasavin}, \citenamefont {Levi-Belenkova}, \citenamefont {Shalin}, \citenamefont {Ginzburg}, \citenamefont {Markovich},\ and\ \citenamefont {Zayats}}]{ginzburg-opex}%
  \BibitemOpen
  \bibfield  {author} {\bibinfo {author} {\bibfnamefont {D.}~\bibnamefont {Vestler}}, \bibinfo {author} {\bibfnamefont {I.}~\bibnamefont {Shishkin}}, \bibinfo {author} {\bibfnamefont {E.~A.}\ \bibnamefont {Gurvitz}}, \bibinfo {author} {\bibfnamefont {M.~E.}\ \bibnamefont {Nasir}}, \bibinfo {author} {\bibfnamefont {A.}~\bibnamefont {Ben-Moshe}}, \bibinfo {author} {\bibfnamefont {A.~P.}\ \bibnamefont {Slobozhanyuk}}, \bibinfo {author} {\bibfnamefont {A.~V.}\ \bibnamefont {Krasavin}}, \bibinfo {author} {\bibfnamefont {T.}~\bibnamefont {Levi-Belenkova}}, \bibinfo {author} {\bibfnamefont {A.~S.}\ \bibnamefont {Shalin}}, \bibinfo {author} {\bibfnamefont {P.}~\bibnamefont {Ginzburg}}, \bibinfo {author} {\bibfnamefont {G.}~\bibnamefont {Markovich}},\ and\ \bibinfo {author} {\bibfnamefont {A.~V.}\ \bibnamefont {Zayats}},\ }\bibfield  {title} {\enquote {\bibinfo {title} {Circular dichroism enhancement in plasmonic nanorod metamaterials},}\ }\href@noop {} {\bibfield  {journal} {\bibinfo  {journal} {Opt. Express}\
  }\textbf {\bibinfo {volume} {26}},\ \bibinfo {pages} {17841--17848} (\bibinfo {year} {2018})}\BibitemShut {NoStop}%
\bibitem [{\citenamefont {Roth}, \citenamefont {Krasavin},\ and\ \citenamefont {Zayats}(2024)}]{Roth2024}%
  \BibitemOpen
  \bibfield  {author} {\bibinfo {author} {\bibfnamefont {D.~J.}\ \bibnamefont {Roth}}, \bibinfo {author} {\bibfnamefont {A.~V.}\ \bibnamefont {Krasavin}},\ and\ \bibinfo {author} {\bibfnamefont {A.~V.}\ \bibnamefont {Zayats}},\ }\bibfield  {title} {\enquote {\bibinfo {title} {Nanophotonics with plasmonic nanorod metamaterials},}\ }\href@noop {} {\bibfield  {journal} {\bibinfo  {journal} {Laser \& Photonics Reviews}\ }\textbf {\bibinfo {volume} {18}},\ \bibinfo {pages} {2300886} (\bibinfo {year} {2024})}\BibitemShut {NoStop}%
\bibitem [{\citenamefont {Aita}\ \emph {et~al.}(2024{\natexlab{a}})\citenamefont {Aita}, \citenamefont {Roth}, \citenamefont {Zaleska}, \citenamefont {Krasavin}, \citenamefont {Nicholls}, \citenamefont {Shevchenko}, \citenamefont {Rodríguez-Fortuño},\ and\ \citenamefont {Zayats}}]{aita2024radazi}%
  \BibitemOpen
  \bibfield  {author} {\bibinfo {author} {\bibfnamefont {V.}~\bibnamefont {Aita}}, \bibinfo {author} {\bibfnamefont {D.~J.}\ \bibnamefont {Roth}}, \bibinfo {author} {\bibfnamefont {A.}~\bibnamefont {Zaleska}}, \bibinfo {author} {\bibfnamefont {A.~V.}\ \bibnamefont {Krasavin}}, \bibinfo {author} {\bibfnamefont {L.~H.}\ \bibnamefont {Nicholls}}, \bibinfo {author} {\bibfnamefont {M.}~\bibnamefont {Shevchenko}}, \bibinfo {author} {\bibfnamefont {F.}~\bibnamefont {Rodríguez-Fortuño}},\ and\ \bibinfo {author} {\bibfnamefont {A.~V.}\ \bibnamefont {Zayats}},\ }\href {https://arxiv.org/abs/2410.22958} {\enquote {\bibinfo {title} {Longitudinal field controls vector vortex beams in anisotropic epsilon-near-zero metamaterials},}\ } (\bibinfo {year} {2024}{\natexlab{a}}),\ \Eprint {https://arxiv.org/abs/2410.22958} {arXiv:2410.22958 [physics.optics]} \BibitemShut {NoStop}%
\bibitem [{\citenamefont {Milione}\ \emph {et~al.}(2012)\citenamefont {Milione}, \citenamefont {Evans}, \citenamefont {Nolan},\ and\ \citenamefont {Alfano}}]{Milione2012}%
  \BibitemOpen
  \bibfield  {author} {\bibinfo {author} {\bibfnamefont {G.}~\bibnamefont {Milione}}, \bibinfo {author} {\bibfnamefont {S.}~\bibnamefont {Evans}}, \bibinfo {author} {\bibfnamefont {D.~A.}\ \bibnamefont {Nolan}},\ and\ \bibinfo {author} {\bibfnamefont {R.~R.}\ \bibnamefont {Alfano}},\ }\bibfield  {title} {\enquote {\bibinfo {title} {Higher order pancharatnam-berry phase and the angular momentum of light},}\ }\href@noop {} {\bibfield  {journal} {\bibinfo  {journal} {Phys. Rev. Lett.}\ }\textbf {\bibinfo {volume} {108}},\ \bibinfo {pages} {190401} (\bibinfo {year} {2012})}\BibitemShut {NoStop}%
\bibitem [{\citenamefont {Aita}\ \emph {et~al.}(2024{\natexlab{b}})\citenamefont {Aita}, \citenamefont {Shevchenko}, \citenamefont {Rodr\'{\i}guez-Fortu\~no},\ and\ \citenamefont {Zayats}}]{aita2024PRB}%
  \BibitemOpen
  \bibfield  {author} {\bibinfo {author} {\bibfnamefont {V.}~\bibnamefont {Aita}}, \bibinfo {author} {\bibfnamefont {M.}~\bibnamefont {Shevchenko}}, \bibinfo {author} {\bibfnamefont {F.~J.}\ \bibnamefont {Rodr\'{\i}guez-Fortu\~no}},\ and\ \bibinfo {author} {\bibfnamefont {A.~V.}\ \bibnamefont {Zayats}},\ }\bibfield  {title} {\enquote {\bibinfo {title} {Propagation of focused scalar and vector vortex beams in anisotropic media: A semianalytical approach},}\ }\href@noop {} {\bibfield  {journal} {\bibinfo  {journal} {Physical Review B}\ }\textbf {\bibinfo {volume} {109}},\ \bibinfo {pages} {125433} (\bibinfo {year} {2024}{\natexlab{b}})}\BibitemShut {NoStop}%
\bibitem [{\citenamefont {Gao}\ \emph {et~al.}(2020)\citenamefont {Gao}, \citenamefont {Speirits}, \citenamefont {Castellucci}, \citenamefont {Franke-Arnold}, \citenamefont {Barnett},\ and\ \citenamefont {G\"otte}}]{Gao2020}%
  \BibitemOpen
  \bibfield  {author} {\bibinfo {author} {\bibfnamefont {S.}~\bibnamefont {Gao}}, \bibinfo {author} {\bibfnamefont {F.~C.}\ \bibnamefont {Speirits}}, \bibinfo {author} {\bibfnamefont {F.}~\bibnamefont {Castellucci}}, \bibinfo {author} {\bibfnamefont {S.}~\bibnamefont {Franke-Arnold}}, \bibinfo {author} {\bibfnamefont {S.~M.}\ \bibnamefont {Barnett}},\ and\ \bibinfo {author} {\bibfnamefont {J.~B.}\ \bibnamefont {G\"otte}},\ }\bibfield  {title} {\enquote {\bibinfo {title} {Paraxial skyrmionic beams},}\ }\href@noop {} {\bibfield  {journal} {\bibinfo  {journal} {Phys. Rev. A}\ }\textbf {\bibinfo {volume} {102}},\ \bibinfo {pages} {053513} (\bibinfo {year} {2020})}\BibitemShut {NoStop}%
\bibitem [{\citenamefont {Kr\'{o}l}\ \emph {et~al.}(2021)\citenamefont {Kr\'{o}l}, \citenamefont {Sigurdsson}, \citenamefont {Rechci\'{n}ska}, \citenamefont {Oliwa}, \citenamefont {Tyszka}, \citenamefont {Bardyszewski}, \citenamefont {Opala}, \citenamefont {Matuszewski}, \citenamefont {Morawiak}, \citenamefont {Mazur}, \citenamefont {Piecek}, \citenamefont {Kula}, \citenamefont {Lagoudakis}, \citenamefont {Piętka},\ and\ \citenamefont {Szczytko}}]{Krol2021}%
  \BibitemOpen
  \bibfield  {author} {\bibinfo {author} {\bibfnamefont {M.}~\bibnamefont {Kr\'{o}l}}, \bibinfo {author} {\bibfnamefont {H.}~\bibnamefont {Sigurdsson}}, \bibinfo {author} {\bibfnamefont {K.}~\bibnamefont {Rechci\'{n}ska}}, \bibinfo {author} {\bibfnamefont {P.}~\bibnamefont {Oliwa}}, \bibinfo {author} {\bibfnamefont {K.}~\bibnamefont {Tyszka}}, \bibinfo {author} {\bibfnamefont {W.}~\bibnamefont {Bardyszewski}}, \bibinfo {author} {\bibfnamefont {A.}~\bibnamefont {Opala}}, \bibinfo {author} {\bibfnamefont {M.}~\bibnamefont {Matuszewski}}, \bibinfo {author} {\bibfnamefont {P.}~\bibnamefont {Morawiak}}, \bibinfo {author} {\bibfnamefont {R.}~\bibnamefont {Mazur}}, \bibinfo {author} {\bibfnamefont {W.}~\bibnamefont {Piecek}}, \bibinfo {author} {\bibfnamefont {P.}~\bibnamefont {Kula}}, \bibinfo {author} {\bibfnamefont {P.~G.}\ \bibnamefont {Lagoudakis}}, \bibinfo {author} {\bibfnamefont {B.}~\bibnamefont {Piętka}},\ and\ \bibinfo {author} {\bibfnamefont {J.}~\bibnamefont {Szczytko}},\ }\bibfield  {title} {\enquote
  {\bibinfo {title} {Observation of second-order meron polarization textures in optical microcavities},}\ }\href@noop {} {\bibfield  {journal} {\bibinfo  {journal} {Optica}\ }\textbf {\bibinfo {volume} {8}},\ \bibinfo {pages} {255--261} (\bibinfo {year} {2021})}\BibitemShut {NoStop}%
\bibitem [{\citenamefont {Shen}(2021)}]{Shen2021}%
  \BibitemOpen
  \bibfield  {author} {\bibinfo {author} {\bibfnamefont {Y.}~\bibnamefont {Shen}},\ }\bibfield  {title} {\enquote {\bibinfo {title} {Topological bimeronic beams},}\ }\href {https://doi.org/10.1364/OL.431122} {\bibfield  {journal} {\bibinfo  {journal} {Opt. Lett.}\ }\textbf {\bibinfo {volume} {46}},\ \bibinfo {pages} {3737--3740} (\bibinfo {year} {2021})}\BibitemShut {NoStop}%
\bibitem [{\citenamefont {Dreher}\ \emph {et~al.}(2024)\citenamefont {Dreher}, \citenamefont {Neuhaus}, \citenamefont {Janoschka}, \citenamefont {R{\"o}dl}, \citenamefont {Meiler}, \citenamefont {Frank}, \citenamefont {Davis}, \citenamefont {Giessen},\ and\ \citenamefont {zu~Heringdorf}}]{Dreher2024}%
  \BibitemOpen
  \bibfield  {author} {\bibinfo {author} {\bibfnamefont {P.}~\bibnamefont {Dreher}}, \bibinfo {author} {\bibfnamefont {A.}~\bibnamefont {Neuhaus}}, \bibinfo {author} {\bibfnamefont {D.}~\bibnamefont {Janoschka}}, \bibinfo {author} {\bibfnamefont {A.}~\bibnamefont {R{\"o}dl}}, \bibinfo {author} {\bibfnamefont {T.~C.}\ \bibnamefont {Meiler}}, \bibinfo {author} {\bibfnamefont {B.}~\bibnamefont {Frank}}, \bibinfo {author} {\bibfnamefont {T.~J.}\ \bibnamefont {Davis}}, \bibinfo {author} {\bibfnamefont {H.}~\bibnamefont {Giessen}},\ and\ \bibinfo {author} {\bibfnamefont {F.~M.}\ \bibnamefont {zu~Heringdorf}},\ }\bibfield  {title} {\enquote {\bibinfo {title} {{Spatiotemporal topology of plasmonic spin meron pairs revealed by polarimetric photo-emission microscopy}},}\ }\href {https://doi.org/10.1117/1.AP.6.6.066007} {\bibfield  {journal} {\bibinfo  {journal} {Advanced Photonics}\ }\textbf {\bibinfo {volume} {6}},\ \bibinfo {pages} {066007} (\bibinfo {year} {2024})}\BibitemShut {NoStop}%
\bibitem [{\citenamefont {Afanasev}\ \emph {et~al.}(2023)\citenamefont {Afanasev}, \citenamefont {Kingsley-Smith}, \citenamefont {Rodr{\'i}guez-Fortu{\~n}o},\ and\ \citenamefont {Zayats}}]{afanasiev-APN}%
  \BibitemOpen
  \bibfield  {author} {\bibinfo {author} {\bibfnamefont {A.}~\bibnamefont {Afanasev}}, \bibinfo {author} {\bibfnamefont {J.}~\bibnamefont {Kingsley-Smith}}, \bibinfo {author} {\bibfnamefont {F.~J.}\ \bibnamefont {Rodr{\'i}guez-Fortu{\~n}o}},\ and\ \bibinfo {author} {\bibfnamefont {A.~V.}\ \bibnamefont {Zayats}},\ }\bibfield  {title} {\enquote {\bibinfo {title} {{Nondiffractive three-dimensional polarization features of optical vortex beams}},}\ }\href@noop {} {\bibfield  {journal} {\bibinfo  {journal} {Advanced Photonics Nexus}\ }\textbf {\bibinfo {volume} {2}},\ \bibinfo {pages} {026001} (\bibinfo {year} {2023})}\BibitemShut {NoStop}%
\bibitem [{\citenamefont {Zaleska}\ \emph {et~al.}(2024)\citenamefont {Zaleska}, \citenamefont {Krasavin}, \citenamefont {Zayats},\ and\ \citenamefont {Dickson}}]{zaleska2024copper}%
  \BibitemOpen
  \bibfield  {author} {\bibinfo {author} {\bibfnamefont {A.}~\bibnamefont {Zaleska}}, \bibinfo {author} {\bibfnamefont {A.~V.}\ \bibnamefont {Krasavin}}, \bibinfo {author} {\bibfnamefont {A.~V.}\ \bibnamefont {Zayats}},\ and\ \bibinfo {author} {\bibfnamefont {W.}~\bibnamefont {Dickson}},\ }\bibfield  {title} {\enquote {\bibinfo {title} {Copper-based core--shell metamaterials with ultra-broadband and reversible enz tunability},}\ }\href@noop {} {\bibfield  {journal} {\bibinfo  {journal} {Materials Advances}\ } (\bibinfo {year} {2024})}\BibitemShut {NoStop}%
\bibitem [{\citenamefont {Johnson}\ and\ \citenamefont {Christy}(1972)}]{JohnsonChristy1972}%
  \BibitemOpen
  \bibfield  {author} {\bibinfo {author} {\bibfnamefont {P.~B.}\ \bibnamefont {Johnson}}\ and\ \bibinfo {author} {\bibfnamefont {R.~W.}\ \bibnamefont {Christy}},\ }\bibfield  {title} {\enquote {\bibinfo {title} {Optical constants of the noble metals},}\ }\href {https://doi.org/10.1103/PhysRevB.6.4370} {\bibfield  {journal} {\bibinfo  {journal} {Phys. Rev. B}\ }\textbf {\bibinfo {volume} {6}},\ \bibinfo {pages} {4370--4379} (\bibinfo {year} {1972})}\BibitemShut {NoStop}%
\bibitem [{\citenamefont {Wu}\ \emph {et~al.}(2023)\citenamefont {Wu}, \citenamefont {Bykov}, \citenamefont {Krasavin}, \citenamefont {Nasir},\ and\ \citenamefont {Zayats}}]{jingyi-apl}%
  \BibitemOpen
  \bibfield  {author} {\bibinfo {author} {\bibfnamefont {J.}~\bibnamefont {Wu}}, \bibinfo {author} {\bibfnamefont {A.~Y.}\ \bibnamefont {Bykov}}, \bibinfo {author} {\bibfnamefont {A.~V.}\ \bibnamefont {Krasavin}}, \bibinfo {author} {\bibfnamefont {M.~E.}\ \bibnamefont {Nasir}},\ and\ \bibinfo {author} {\bibfnamefont {A.~V.}\ \bibnamefont {Zayats}},\ }\bibfield  {title} {\enquote {\bibinfo {title} {Thermal control of polarization of light with nonlocal plasmonic anisotropic metamaterials},}\ }\href@noop {} {\bibfield  {journal} {\bibinfo  {journal} {Applied Physics Letters}\ }\textbf {\bibinfo {volume} {123}},\ \bibinfo {pages} {171701} (\bibinfo {year} {2023})}\BibitemShut {NoStop}%
\end{thebibliography}%
\clearpage

\newpage 
\begin{center}
\large{\bf Supporting Information\\  Polarisation conversion and optical meron topologies in anisotropic epsilon-near-zero metamaterials}\\
\medskip
\large{Vittorio Aita, Anastasiia Zaleska, Henry J. Putley and Anatoly V. Zayats}\\

\end{center}
\renewcommand{\thepage}{S\arabic{page}} 
\setcounter{page}{1}
\renewcommand{\figurename}{Supplementary Figure}
\setcounter{figure}{0}  
\renewcommand{\thesection}{S\arabic{section}}  
\setcounter{section}{0}

\begin{figure}[!h]
    \centering
    \includegraphics[width=0.9\textwidth]{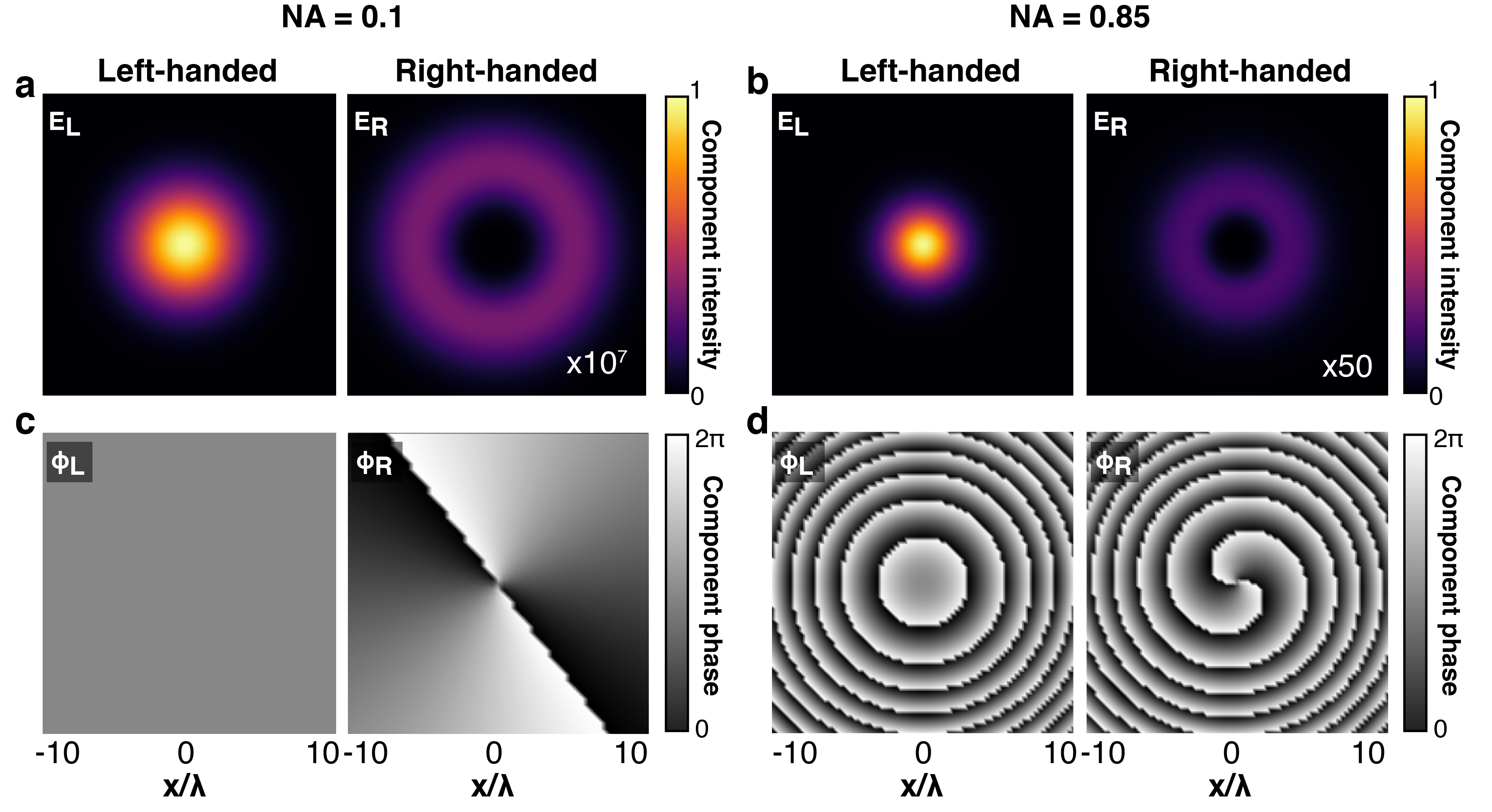}
    \caption{\textbf{Vortex generation in the metamaterial by circularly polarised light.} A left-handed (L) circularly polarised beam propagates through the metamaterial at a wavelength $\lambda \approx$800~nm, red-shifted from the ENZ regime in (a,c) weak (NA = 0.1) and (b,d) strong (NA = 0.85) regime. The transmitted light is decomposed onto its (R) right-and (L) left-handed circularly polarised components as indicated by the labels in each panel. (a,b) Intensity of the R and L components as indicated by the labels. The colour scale is the same for each pair of amplitude plots, to allow direct comparison at fixed NA. The R amplitude has been scaled by a factor shown in the respective panels. (c,d) Phase distribution of the R and L components, wrapped in [0, 2$\pi$]. The effect of the tight focusing is visible as the appearance of concentric rings. The emergence of a topological charge 2 is seen as the line splitting the phase through the centre, only shown by the R components.}
    \label{fig:si_circ_vortex_comps}
\end{figure}

\begin{figure}[!h]
\centering
    \includegraphics[width = \textwidth]{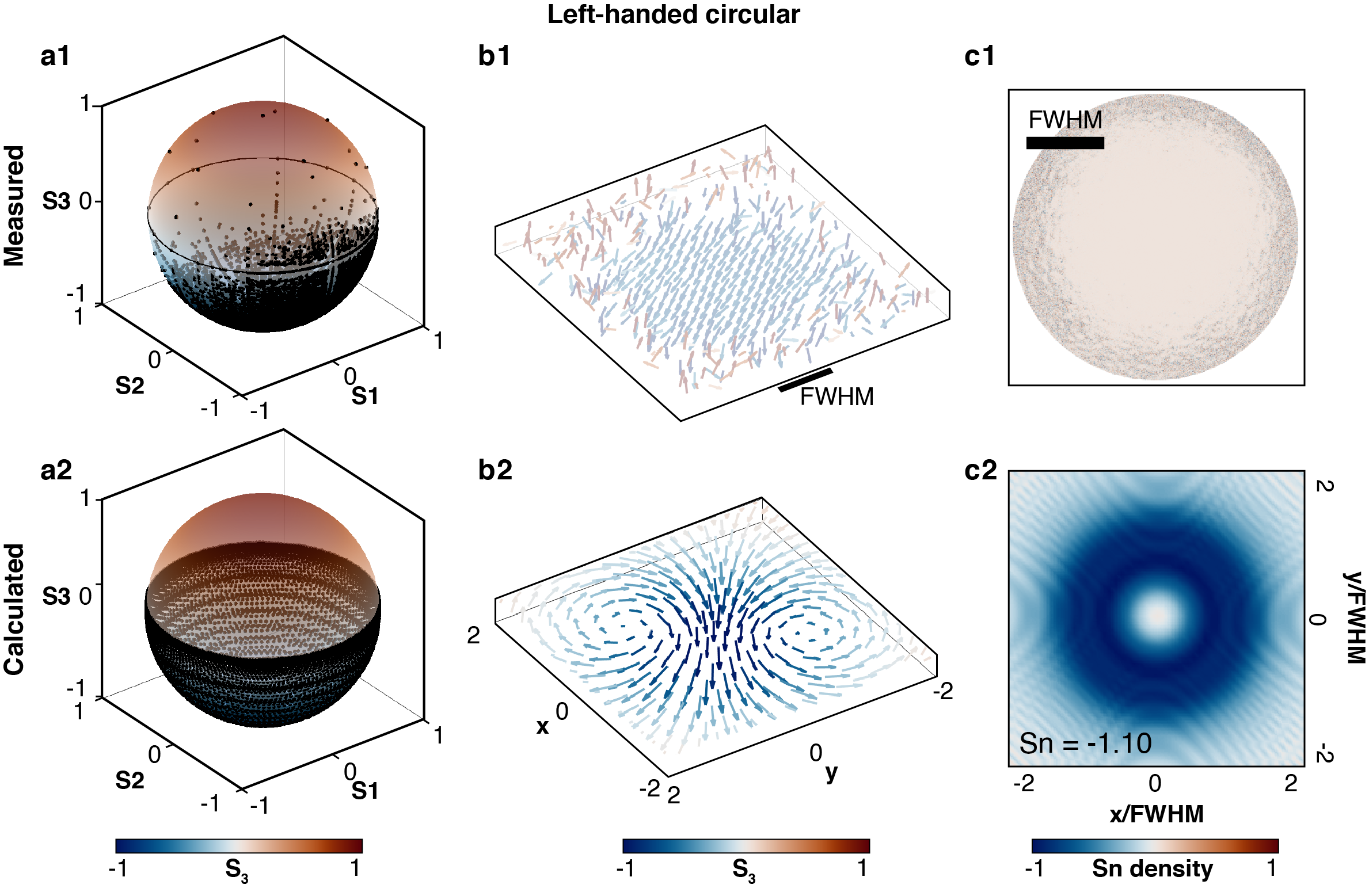}
    \caption{\textbf{Additional data for LHC.} Supporting data to Fig.~\ref{fig:off_res} showing (a) experimental and (b) theoretical results for a left-handed circularly polarised tightly focused Gaussian beam, propagating through the metamaterial for its hyperbolic dispersion. Analysing the state of polarisation obtained in output, the following information is retrieved: (1) coverage of the Poincaré sphere, (2) three-dimensional distribution of the vector field $\vb{\Sigma}$ and (3) the skyrmion number density.}
    \label{fig:si_LHCpoin}
\end{figure}

\begin{figure}[!h]
\centering
    \includegraphics[width = \textwidth]{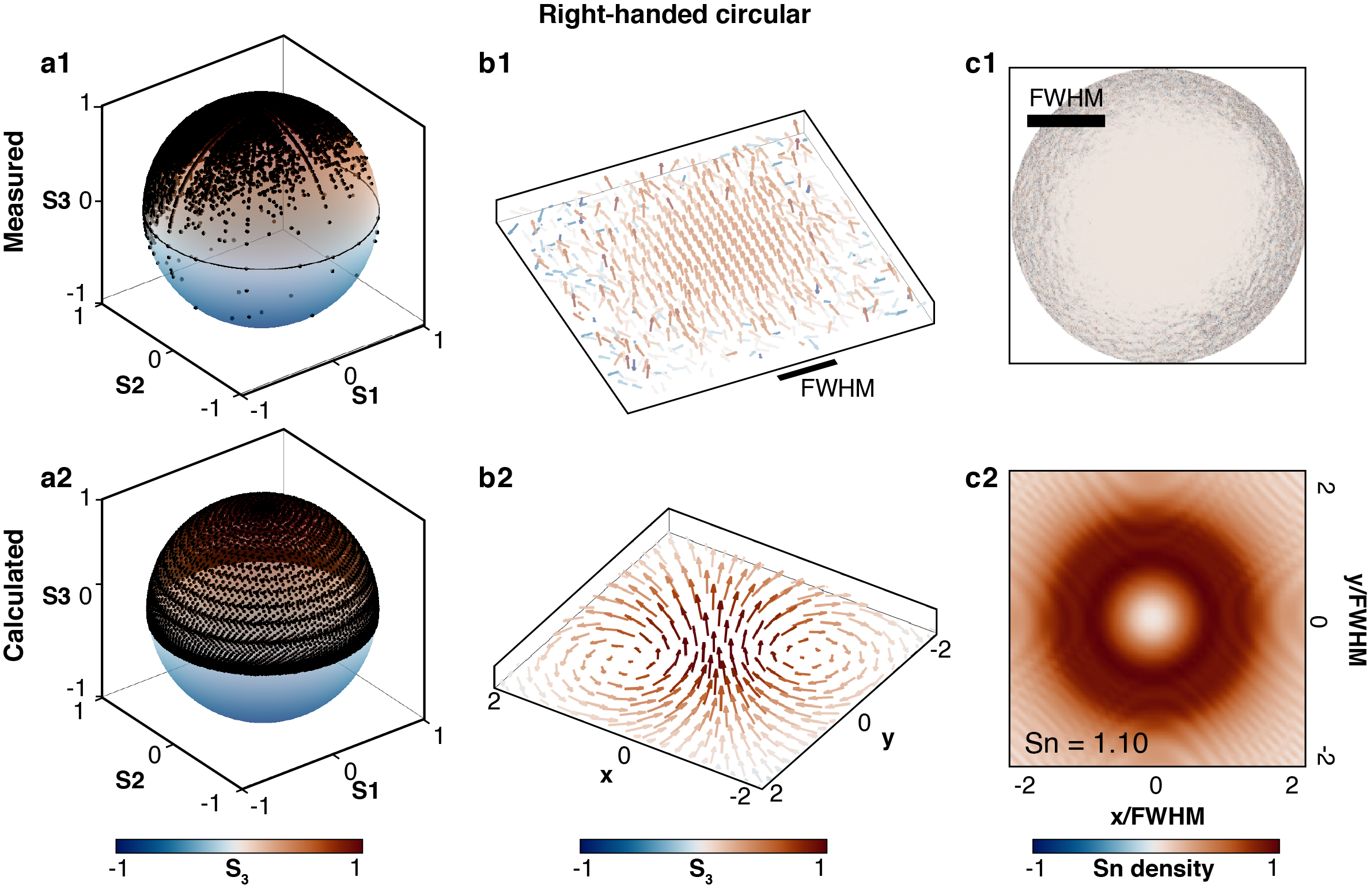}
    \caption{\textbf{Additional data for RHC.} Results analogous to Fig.~\ref{fig:si_LHCpoin}, obtained for right-handed circular polarisation.}
    \label{fig:si_RHCpoin}
\end{figure}

\begin{figure}[!h]
\centering
    \includegraphics[width = \textwidth]{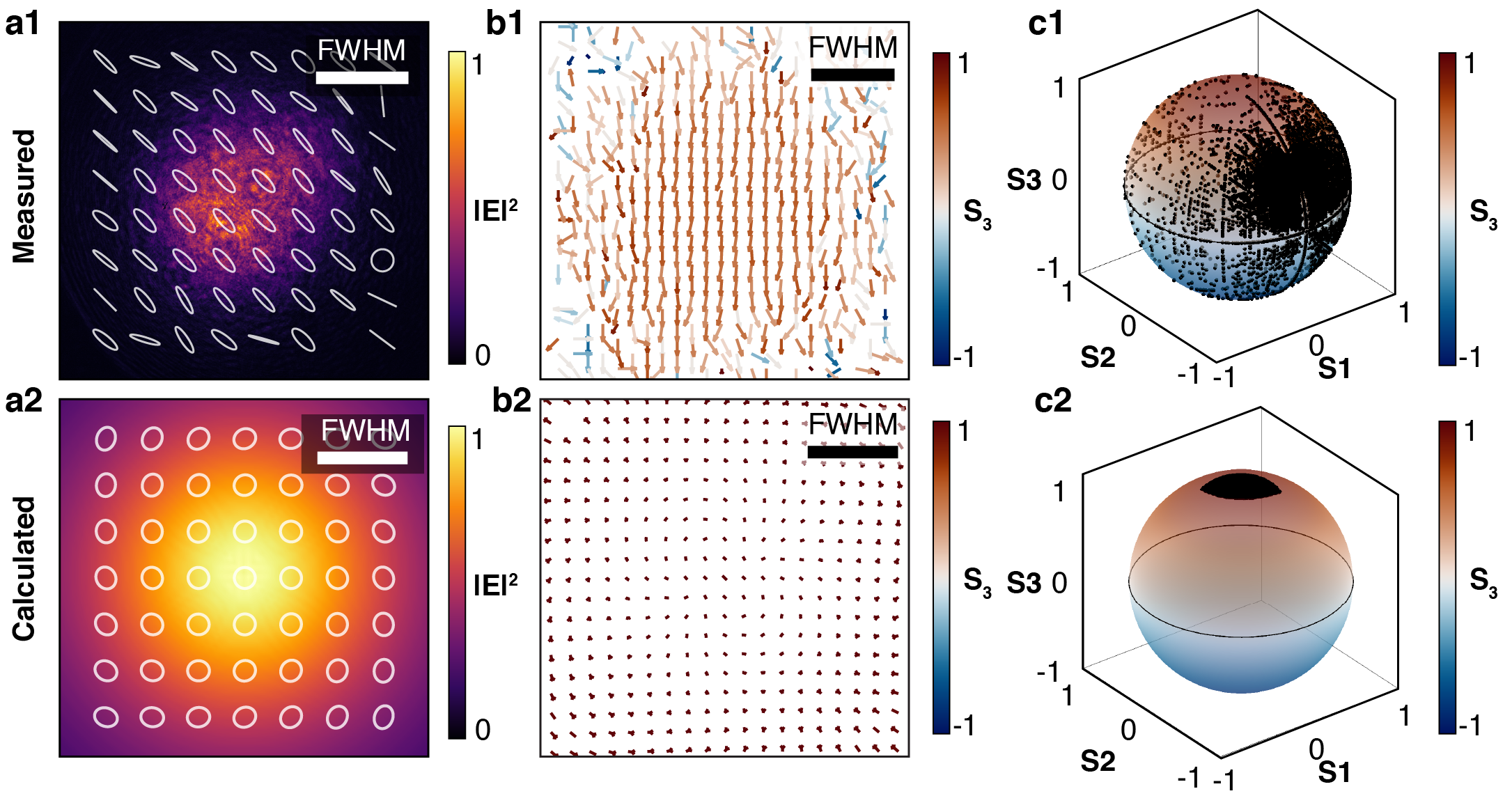}
    \caption{\textbf{Propagation in the elliptic dispersion regime.} Analogous results to Fig.~\ref{fig:off_res} obtained for a right-handed circularly polarised Gaussian beam and a wavelength of $\lambda =$ 488~nm. (1) Experimental and (2) theoretical results for (a) the intensity profile and the state of polarisation, (b) the spatial distribution of the corresponding vector field $\vb{\Sigma}$ and (c) the Poincaré sphere coverage.}
    \label{fig:si_blue_shift}
\end{figure}

\begin{figure}[!h]
\centering
    \includegraphics[width = \textwidth]{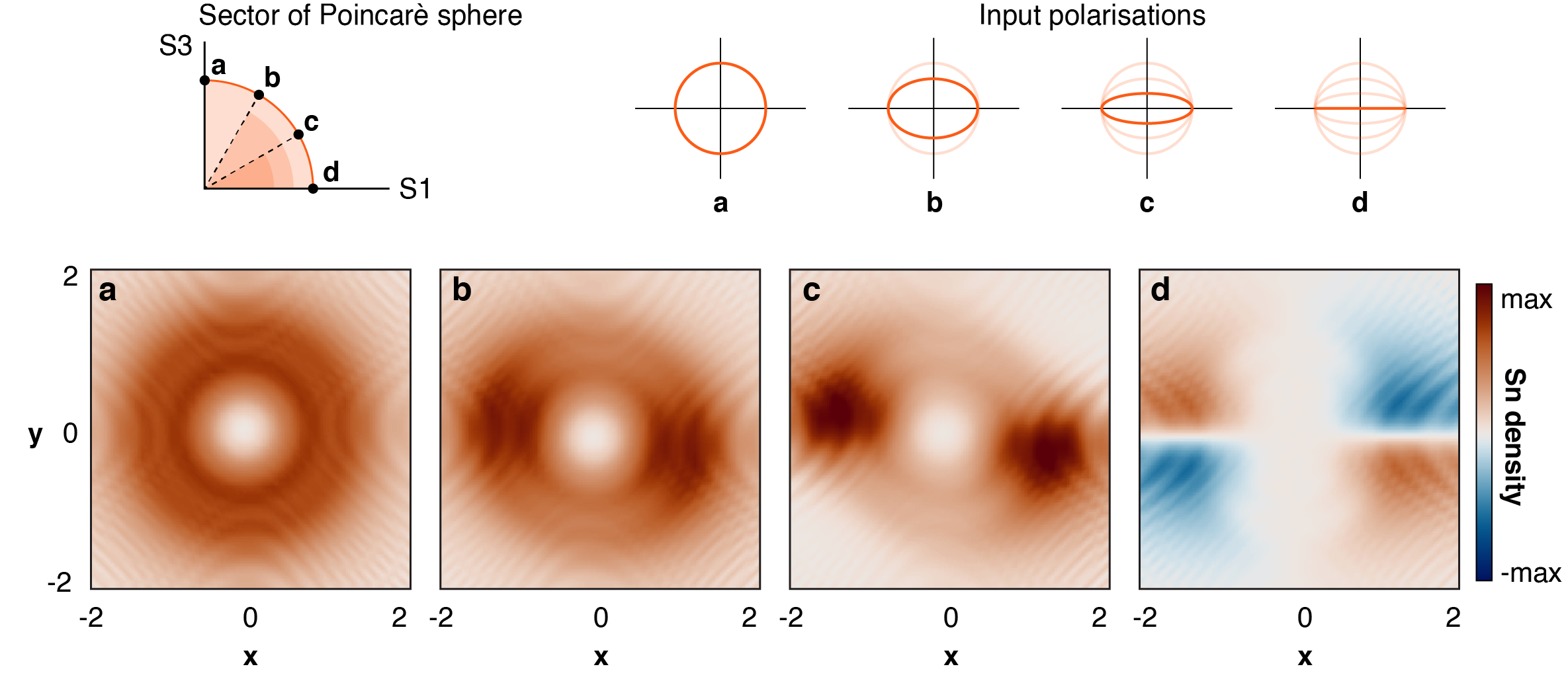}
    \caption{\textbf{Dependence of skyrmion number density on input ellipticity.} Data in support of Fig.~\ref{fig:imperfections}, showing the evolution the skyrmion number density as the input beam ellipticity is reduced from (a) circular to (d) linear polarisation.}
    \label{fig:si_ellipticity}
\end{figure}

\end{document}